\documentclass[11pt]{article}
\usepackage{graphicx}
\def\title{\begin{center}\Large\bf}
\def\author(s){\vspace{0.3cm}\large\rm}
\def\text{\end{center}}

\begin{document}


\title
Gamma-Ray Bursts as Manifestation of Collisions of Primordial Black
Holes with Stars

\bigskip



\author(s)
B.E. Zhilyaev$^{\rm \,1,\, 2}$

\bigskip

\smallskip

\noindent $^1${\small {\it Main Astronomical Observatory, NAS of
Ukraine,  27 Zabolotnoho, 03680 Kiev, Ukraine}} \\
\noindent {\small {\it e-mail:}} {\small{\bf zhilyaev@mao.kiev.ua}}

\smallskip

\noindent $^2${\small {\it International Centre for Astronomical,
Medical and Ecological Research \\ Terskol settlement,
Kabardino-Balkaria, 361605 Russia}} \\


\text


\section*{Abstract}

The gamma-ray burst phenomenon is among the most powerful transient
in the Universe. A crucial question appears: whether GRBs include a
uniform population or are separated into different classes? The
gamma-ray spectra of the great majority of the observed GRBs are
definitely non-thermal. Using the BATSE survey data I find that
quite small fraction of GRBs seems to emit the radiation similar to
thermal bremsstrahlung in the range 20 to 300 $keV$. This subclass,
numbering 37 sources, forms precisely allocated cluster on the
two-color diagram and contains no more than 2\% of all the bursts
detected. I suggest that these bursts may perhaps occur from
collision of stars with primordial black holes (PBH). These objects
are relic of a hot matter in the early Universe. Hypothetically, PBH
had formed due to the collapse during the radiation era before $\sim
10^{-4}$ s since the beginning of the Universe. They could grow up
to horizon mass of about $1 M_{\odot}$. PBH in the vicinity of stars
may be found in consequence of incorporation processes during the
formation of stars from interstellar clouds. At present they can
form the gravitationally captured haloes around stars like the
family of solar comets. The comet paradigm has been used to
understand various aspects of PBH. Their behavior, in terms of
celestial mechanics, may be similar to that seen in solar comets.
Like comets they can undergo a complex orbital evolution process,
driven by secular resonances,  and by a sequence of close encounters
with planets. Eventually, they can collide with the central star or
can be ejected from the star system. Comet collisions with the Sun
and planets are ordinary events in solar system history. On the
analogy, one can support the view that PBH collisions with the
parent star may be quite frequent events in its history, too. PBHs
are the engines driving gamma-ray bursts when collide with the
stars. Entering a stellar atmosphere, PBH is supposed to produce the
gamma-ray burst due to accretion with a duration from a few tenths
of second to a few seconds. It can exhibit the main qualitative
features of GRBs. If the claimed PBH luminosity is no more than the
Eddington limit, a thermal bremsstrahlung model fits leads to the
temperature estimates in the range from tens to hundreds $keV$ and
to the power estimate of $10^{\,35\div 36} \, erg \, s^{-1}$. Their
masses are estimated in the range from thousandths to hundredths of
the solar mass. I found that these burst sources are isotropically
distributed on the sky and are seen from a distance up to 50
\textit{ps}. In this context one may expect that some short GRBs are
observable signatures of primordial black holes in the Universe.

\vspace{1.0cm}

\noindent {{\bf keywords}$\,\,\,\,$\large gamma-rays: bursts --
black hole physics -- (stars:)planetary systems}

\large

\section{INTRODUCTION}

Gamma-ray bursts are the most powerful transient phenomena in the
Universe. The BATSE survey data show that bursts are isotropically
distributed. The gamma-ray burst durations range from milliseconds
to hundreds of seconds, with a bimodal distribution of long bursts
with $\Delta t \geq  2\, s$ and short bursts with $\Delta t \leq 2\,
s$. The most plausible GRB progenitors suggested so far are expected
to lead to a system with a central BH, NS-NS or NS-BH mergers, white
dwarf - black hole mergers, hypernova or collapsars, and
accretion-induced collapse. As mentioned by Meszaros (1999), the
overall energetics from these various progenitors does not differ by
more than about one order of magnitude. The measured gamma-ray
fluences imply a total energy of order $10^{54}\, erg$ (isotropic)
for GRB of cosmological origin, as results from high redshift
observations. This is of the order of the binding energy of one
solar rest mass. Note that there are only two dozen GRBs for which
an estimate of the redshift is available. These  have estimated
either from absorption/emission lines in the optical afterglow or of
host galaxies, or $Fe$ emission lines in the afterglow X-ray
spectra. Thus, only little more than 1\% of GRBs have reliable
cosmological origin, as results from observations.

Different nature of long and short GRB follows from: (1) their
duration distribution with the mean duration of $\sim 20$ sec and
$\sim 0.3$ sec, respectively (Kouveliotou et al. 1993); (2) their
different temporal properties, e.g. number and width of pulses in
the light curve; (3) distributions of their gamma-ray spectra in the
range $\sim 30-1800 \,\, keV$ (Ghirlanda et al. 2003); (4) and,
rather probably, the lack of any afterglow for short GRBs. Notice,
only about 50\% of the bursts display radio or optical counterparts.
No redshift of short GRBs has been measured so far. These sources
have not astronomical counterparts at other wavebands, too. Thus,
the question of  their cosmological origin remains controversial.

Current models, such as the fireball shock model, the simple
standard afterglow model, post-standard afterglow models, the
relativistic blast wave model of gamma-ray bursts provide an
interpretation of the major features of these objects, reproduce the
properties of their light curves, the afterglows in X-ray, optical
and radio, gravitational radiation from progenitors (Meszaros, 1999,
2002; Kobayashi, Meszaros, 2003). At the same time, there still
remain a number of mysteries, concerning progenitors, the nature of
triggering mechanism, the transport of the energy, etc.

In empirical studies GRBs have been subjected to straightforward
statistical analysis to examine a crucial question of whether GRBs
include a uniform population or are separated into different classes
(Fishman, 1995; Mukherjee et al. 1998). Such an approach has the
large heuristic force, but, unfortunately,  had solved none physical
tasks.


\section{The GRB sample and cluster analysis}
We extract our data from the online database
www.batse.msfc.nasa.gov/data/grb/
catalog, which provides many
properties of each burst from the BATSE 4b catalog. We consider
three fluences, F1-F3, in the 20-50, 50-100, and 100-300 $keV$
spectral channels, respectively, and three  colors derived from the
ratios of fluences, $C12 = F1/F2$, $C13 = F1/F3$, and $C23 = F2/F3$.
We make use of the  logical condition $C9 = C12 >1\,\, \& \,\,C23
>1$ to select a subclass of bursts with a particular spectral
density function. One clearly sees from Equation (1) that the above
logical condition picks out sources, noted by C9, with spectrum like
the thermal deceleration radiation spectrum (bremsstrahlung). The
subsequent analysis has quite proved validity of this assumption.
The color-color plot of a subclass C9 is overlaid on a plot of all
bursts from the BATSE 4b catalog. The bifurcation of the sample into
two classes is easily seen in Fig 1 (left). This affords ground for
considering GRBs of a subclass C9 in the frame of one physical
model. The general idea of present work is that these GRB events may
occur from collision of stars with primordial black holes.

PBHs have been claimed to be either interstellar objects suffering
random collisions with chance stars at their way or members of the
family like solar comets orbiting around parent stars.  Entering a
stellar atmosphere, PBH is supposed to produce the gamma-ray burst
due to accretion. Its characteristics can be determined from a
study of the GRB properties such as their fluences, colors,
durations, etc.

The question of burst repetition remains controversial. Strohmayer
et al. (1994) found from the  sample of 260 bursts that the level of
uncertainty in BATSE burst locations limits a repeating fraction of
less than $10\div15 \,\%$ repeaters. The tightened inspection of
gamma-ray burst repetition by analyzing the angular power spectrum
of the BATSE 3B catalog of 1122 bursts showed that no more than 2\%
of all observed bursts can be labeled as repeaters (Tegmark et al.
1996). Furthermore, at 95\% confidence, they conclude that the BATSE
data are consistent with no repetition of classical gamma-ray
bursts. Fig 2 from distributions of sources of the C9 class allows
also to confidently conclude that these bursts were observed no more
than once during 9 years of the BATSE mission. Secondly, the
distribution in galactic coordinates shows that GRBs of a subclass
C9 are isotropically distributed, in a statistical sense. Each burst
comes from a random direction; no repeating events have been
detected from precisely the same direction.

\begin{figure*}
\centering
\resizebox{1.0\hsize}{!}{\includegraphics[angle=000]{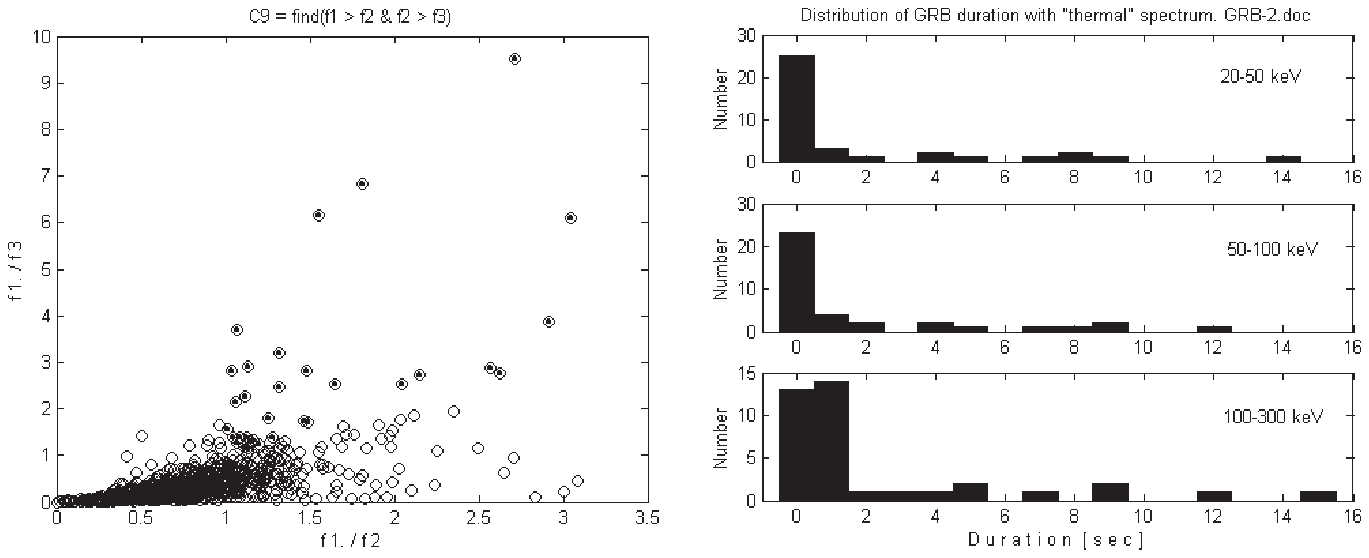}}%
\caption{Left: The colors of GRBs (a subclass C9, filled circles),
defined by the ratio of fluences in BATSE channel 1 (20-50 $keV$) to
channel 2 (50-100 $keV$) vs. channel 1 to channel 3 (100-300 $keV$),
overlaid on a the color-color plot of all bursts detected (open
circles).\newline Right: Histogram bar plots of the gamma-ray burst
durations in seconds relative to the burst trigger time. The
channels 1-3 from top to bottom. About two-thirds of bursts have a
duration about one second, remaining third - up to about twenty
seconds.}
\end{figure*}

\section{Radiation processes with primordial black holes}

A primordial population of black holes is thought was created in
the early Universe. PBH had formed due to the collapse during the
radiation era before $\sim 10^{-4}\, s$ since the beginning of the
Universe. They could grow up to horizon mass of about
$1\,M_{\odot}$. The moment of PBH formation $t_{0}$  depends on
its starting mass (Zeldovich, Novikov, 1966),  $t_{0}(s)\sim
GM/c^{3} \sim 2\cdot 10^{-39}\, M(g)$, where the time is in
seconds, the mass in grams. The hypothesis of PBHs formation near
the cosmological singularity from density and metric fluctuations
validated through numerical calculations by Novikov et al. (1979).
PBH of less than $\sim 10^{15}\,g$ should have evaporated  by now
through the Hawking process. It appears that PBH with mass of
$\sim 10^{15}\,g$ are now the most plentiful. Observations place
an upper limit on an average space density of such PBHs about of
$10^{4}\, ps^{-3}$. But if  PBHs are clustered into galaxies, the
local density can be greater by a factor exceeding $10^{6}$ (Page,
Hawking, 1976; Wright, 1996). This provides an upper limit  of
about $n_{BH} \sim 4\cdot 10^{-46}\,cm^{-3}$ in the Galaxy
(Chapline, 1975; Wright 1996). From this about one PBH we may
expect to find in our solar system on average. This opportunity
was investigated by Zhilyaev (2003) in details.

\begin{figure*}
\centering
\resizebox{0.90\hsize}{!}{\includegraphics[angle=000]{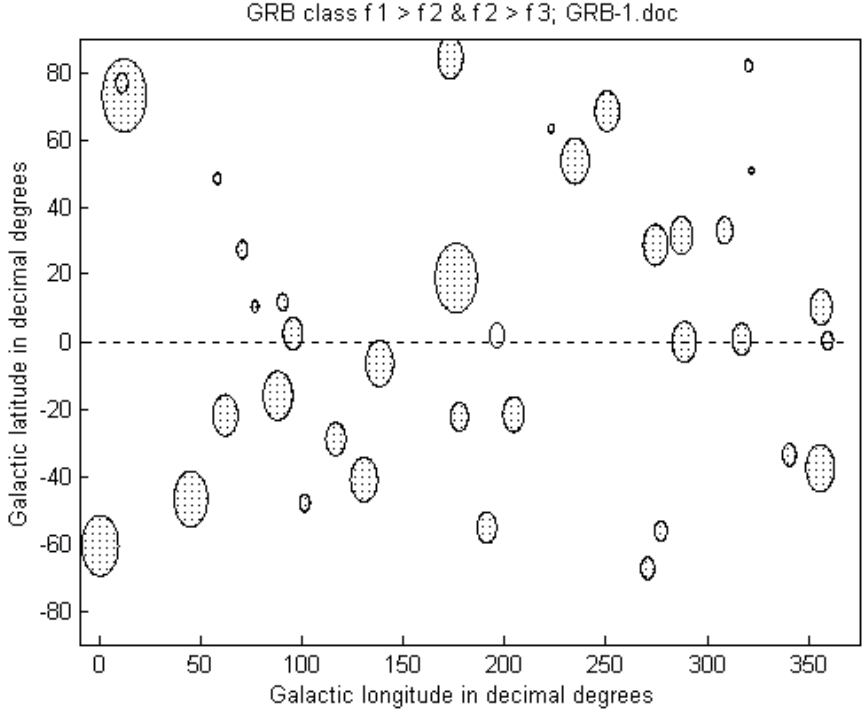}}%
\caption{The distribution in galactic coordinates with positional
error ellipses shows that GRBs of a subclass C9 are isotropically
distributed, in a statistical sense. Bursts come from a random
direction; no coincident events have been detected from just the
same direction. One repeating at the top left  we treated  as a
random coincidence.}
\end{figure*}

Observations of the Hawking radiation from the globular clusters
can provide observational signature of PBHs. Gravitationally
captured PBH haloes around the globular clusters were considered
by Derishev, Belyanin, (1999). EGRET observations of the gamma-ray
luminosity above 100 $MeV$ of five nearby massive globular
clusters placed, however, only the upper limits on the total mass
of PBHs and their mass fraction in these clusters  $\sim 10^{-6}$.
Notice, all the mentioned estimates refer to marginal PBH with
mass of $\sim 10^{15}\, g$. The number density of PBH of greater
mass up to $\sim 1 M_{\odot}$  is one of the most discussing
topics of advanced cosmology. Observations of microlensing events
towards the Large Magellanic Clouds reveals that the event rate is
well above expectation from 'known' stars in the Galactic halo.
The durations of events leads to the lens masses estimate of
roughly $0.3\div 0.7$ solar mass, which is a significant puzzle at
present. Sutherland (1999) noted an extraordinary solution of the
problem, e.g. possibly primordial black holes.

\begin{figure*}
\centering
\resizebox{0.90\hsize}{!}{\includegraphics[angle=000]{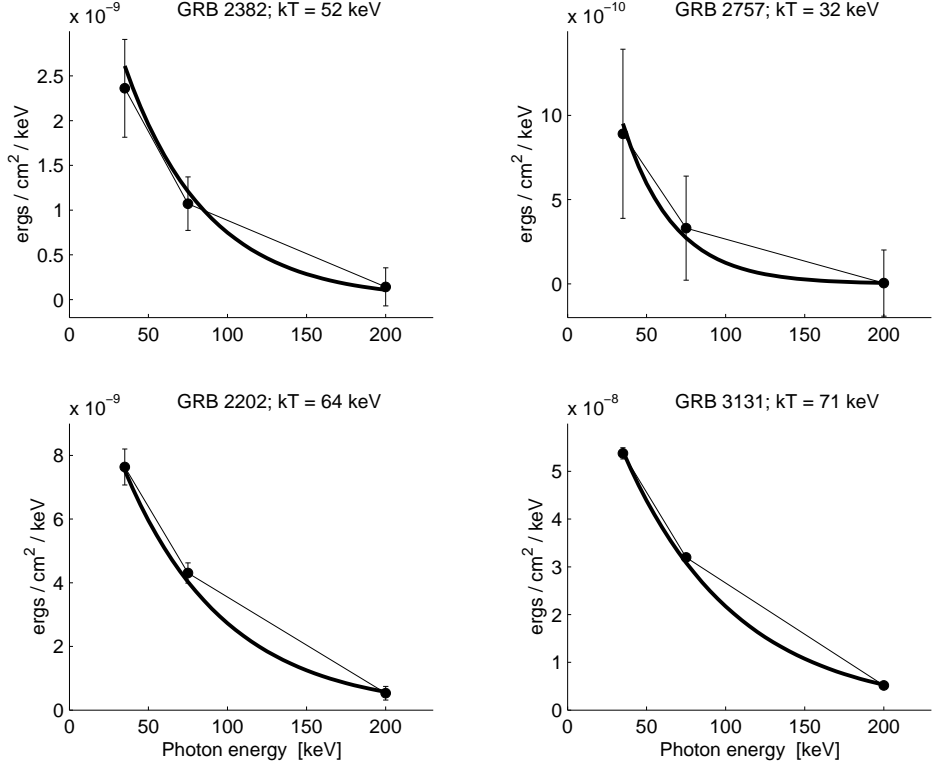}}%
\caption{Some of GRB spectra (all fluences and their errors) and
their thermal bremsstrahlung model fits (heavy lines). Equation (1)
was used to derive fits for the temperature of the bursts.}
\end{figure*}

\subsection{GRBs with the thermal bremsstrahlung spectrum}

In a reference frame, in which the small-mass BH and the target
star are initially at rest, the impact speed is equal to the
escape speed $v_{e}$. To a first approximation this is true for
the BH with nonzero initial speed. So,  for a solar type star the
BH would arrive at the surface with the speed of about $600\, km
\, s^{-1}$, for a giant star its value would be about $350 \,km \,
s^{-1}$ (Allen, 1973). From this a characteristic time $\Delta t$
for GRB sources can be derived, apart a small numerical factor,
$\Delta t \sim  L / v_{e}$, where $L$ is the distance for which
atmospheric attenuation of gamma rays becomes substantial. To
avoid excessive attenuation in stellar photosphere for photons
with energies between $10 \,keV$ and $100 \,MeV$ the burden of
overlying material must amount to $\sim 10\, g\, cm^{-2}$
(Hillier, 1984). For a Sun-like star and a typical giant star
these correspond to altitudes $\sim 300\, km$ and $\sim 7000
\,km$, respectively, above the photosphere level with optical
depth of unity. So, one may expect a characteristic time $\Delta
t$ for GRB sources to be  $0.5$ and $20$ seconds for a Sun-like
star and a giant star, respectively. Note, as follows from Fig 1
(right), histogram plots of  the gamma-ray burst durations of a
subclass C9 display that about two-thirds of bursts have duration
about one second, remaining third - up to about twenty seconds.

The energy spectrum of the radiation from GRB depends on the
characteristics of the source itself. Since the observed gamma-ray
spectrum is below $0.511 \,MeV$, the threshold for the pair
production, the fireball is expected with the luminosity not much
above than Eddington limit. Such a spectrum can produce an optically
thin plasma with thermal bremsstrahlung. The flux leaving the
fireball on account of thermal bremsstrahlung is given by (Allen,
1973; Hillier, 1984)

\begin{equation}
I_{\nu} = 5.44\cdot 10^{-39}\cdot Z^{2}\cdot g N N_{e}/\surd T
\exp(-h\nu/k/T),\,\,\, erg\, cm^{-3}\, s^{-1}\,
sterad^{-1}\,Hz^{-1}
\end{equation}
\smallskip

\noindent where $Z$ is the charge of particles, $g$ is the Gaunt
factor, $N = N_{e}$ is the number density of ions and electrons.
The summarized power of the fireball integrated along the line of
sight  is

\begin{equation}
 I_{\nu} =1.44\cdot 10^{-27}\cdot \surd T \cdot Z^{2}\cdot EM,\,\,\, erg\, s^{-1},
\end{equation}
\smallskip

\noindent where $EM$ is the volume emission measure

\begin{equation}
  EM = \int N N_{e}d V = 0.85\cdot N_{e}^{2}V
\end{equation}
\smallskip

\noindent For cosmic abundances and uniform density $EM =
0.85\cdot N_{e}^{2}V$, where $V$ is the fireball volume (Ness,
2004). As proposed, the fireball extent corresponds to roughly the
run of gamma rays at energies between $\sim 10\, keV$ and $\sim
100\, MeV$ in stellar atmosphere, i.e. the atmospheric depth where
the mass along the line of sight reaches $\sim 10\, g$. Adopting
typical values of $N_{e}$ from Allen (1973), we may evaluate  the
fireball radius for a Sun-like star from $300$ to $1000 \, km$.
The fireball radius in the atmosphere of a typical giant star is
about 20 times larger. For the total optical depth of the fireball
for Thomson scattering we may write

\begin{equation}
\tau \simeq \sigma_{T} N_{e}L
\end{equation}
\smallskip

\noindent where $\sigma_{T}$ is the cross-section for Thomson
scattering. With the values of $N_{e}$ and  $L$ mentioned  above the
optical depth are $0.2\ldots 0.6$ and $\sim 4$ for the Sun and for a
typical giant, respectively. Thus, we can suspect, that the
condition of low optical depth is held approximately true. Fig 3
shows some of GRB spectra and their thermal bremsstrahlung model
fits. Equation (1) can be used to derive fits for the temperature of
the bursts. Equation (2) leads to (isotropic) energy estimate. Thus,
a model fit leads to the temperature estimates in the range $30$ to
$70\, keV$. E.g., in GRB2757 a model fit leads to a temperature
estimate of $T = 32 \pm 3 \,keV$ and to a power estimate of
$1.5\cdot 10^{35}\, \,erg\, s^{-1}$, for a solar type star and by
order of magnitude greater for a giant star. If  the BH luminosity
is taken to be the Eddington limit

\begin{equation}
 L_{c} = 3\cdot 10^{4} L_{\odot} M_{BH} / M_{\odot}
\end{equation}
\smallskip

\noindent then a typical fluence in BATSE channel 1 $(20-50\, keV)$
of  $F1 \sim 5\cdot 10^{-7} erg\, cm^{-2}\,s^{-1}$ should correspond
to the distance of source of about 50 parsecs. This leads to the BH
mass estimate of $\sim 0.002 M_{\odot}$ for a Sun-like star and by
order of magnitude greater for a giant star. Thus, the PBH mass may
amount from some thousandth to some hundredth parts of the solar
mass.


From the burst rate related to a new subclass of GRB discussed
above formal number density of BH can be calculated. Let number
densities are $n_{BH}$ and $n$ for the BH and target stars
respectively. Then the encounter rate of BH per unit volume can be
written:

\begin{equation}
 \Gamma  =  n_{BH}\cdot n\cdot v\cdot \sigma \, , \,\,\,cm^{-3}\, s^{-1}
\end{equation}
\smallskip

\noindent The values of $\lg (n)$ for giants, main sequence stars
and white dwarfs are: -3.2, -1.2, -2.3 $ps^{-3}$, respectively,
according to Allen (1973). By the adoption of the mean values of the
stellar radius  $r$ equal to $10$, $1$ and $0.01$ in solar units for
the above targets the weighted values of $n$ and $r$ are: $n =
0.069\, ps^{-3}$,  $r = 1.01 \, R_{\odot}$. Let adopt the number
density $n \simeq 0.069\, ps^{-3}$, the cross section for closest
encounter $\sigma = \pi \, R_{\odot}^{2}$ and the mean stellar
velocity of $\sim 100\, km\, s^{-1}$. Then for an average rate of
bursts $\Gamma \approx 4\, yr^{-1}$ we can expect $n_{BH}$ of order
$3\, 10^{15} \,ps^{-3}$, assuming uniform distribution. The enormous
amount of this estimate disagrees with the uniform distribution and
forces to conclude that PBH should be clustered around stars forming
the gravitationally binding systems.

\section{PBHs in the light of the comet paradigm}

PBH in the vicinity of stars may be found either in consequence of
captures processes or incorporation during the formation of stars
from interstellar clouds. The binding systems may result, in
particular, from tidal capture and exchange encounters (Johnston \&
Verbunt, 1996). Tidal capture occurs when a PBH transfers some of
its kinetic energy to tides in another star during a close passage,
and enough tidal energy is dissipated to bind the PBH in orbit
around its captor. An exchange encounter occurs when a PBH ejects
one of the stars in a binary in a close encounter, and takes its
place. However, these processes are efficient only with a quite
massive PBH mainly in globular clusters, where the average distance
between stars is relatively small.

The incorporation of PBHs during the formation of stars and other
gravitationally bound objects was analyzed by Derishev, Belyanin
(1999). The detailed description of a gravitational incorporation
requires exact calculations of the collapse dynamics. Two of the
simplest cases were analyzed. These authors argued that in the
free-fall contraction relationship between the PBH number density
and the average one remains constant. PBHs become trapped inside a
protostar. In the case of an adiabatic contraction an appreciable
fraction of PBHs forms the gravitationally captured haloes around
the protostar.

We may use the comet paradigm to understand various aspects of the
PBH phenomenon. As in the solar system, small-mass PBHs orbiting
other stars may resemble solar comets in both their dynamical
history and orbital evolutions over long periods of time. Their
behavior, in terms of celestial mechanics, may be similar to that
seen in solar comets. The idea of  distant reservoirs of comets
known as the Oort cloud, the Edgeworth-Kuiper belt, Centaurs and
Jupiter-family comets may be appreciable to other star systems
with PBH companions. It appears that in terms of orbits the comet
analogue may be appreciable in unmodified form. Stress some
essential features of the comet paradigm. (1) No more than 18
percent of comets have a period less than 20 years. The others do
not follow perfect elliptic or hyperbolic orbits. Their orbital
evolution may be typically chaotic due to gravitational
perturbations by the giant planets. None of comets is coming from
outside the solar system. They form a gravitationally bound halo
around the Sun. (2) Comet collisions with the Sun and planets are
ordinary events in solar system history. For example, the Kreutz
sungrazing group of comets shows extraordinarily small perihelion
distances (Bailey, 1992).  So, the orbit of comet Ikeya-Seki
passed in perihelion at a distance from the center of the Sun of
only 1.67 times its radius. Remember also the collision of  comet
Shoemaker-Levy 9 with Jupiter in July 1994.

On the analogy  of the sungrazing group of comets, one can clearly
see some PBHs  as 'stargrazers' in other stellar systems. Thus, in
terms of orbits we can suppose all these features to be also
inherent gravitationally bound PBH around other stars. All these
appear to support the view that PBH collisions with the parent star
may be quite frequent events in its history. In this context one may
expect that some short GRBs are observable signatures of primordial
black holes in the Universe.

\section{Conclusion}

To summarize, some GRB progenitors can be related to primordial
black holes, which have formed the gravitationally captured haloes
around stars like the family of solar comets. PBHs  are relic of a
hot matter in the early Universe. They could be captured during
the formation of stars from interstellar clouds. PBHs may be
randomly injected from distant reservoirs similar to the Oort
cloud or the Edgeworth-Kuiper belt in the solar system. These
objects can undergo a complex orbital evolution process, driven by
secular resonances,  and by a sequence of close encounters with
planets. Eventually, they can collide with the central star or can
be ejected from the star system. PBHs are the engines driving
gamma-ray bursts when collide with the parent stars. They can
exhibit the main qualitative features of GRBs. Entering a stellar
atmosphere, PBH can produce the gamma-ray burst due to accretion
with a duration from a few tenths of second to a few seconds.
Adopting the Eddington luminosity for PBH, a thermal
bremsstrahlung model fits leads to the temperature estimates in
the range of some tens $keV$ and to the power estimate of
$10^{\,35\div 36}\,\, erg\, s^{-1}$. Their masses are estimated in
the range from thousandths to hundredths of the solar mass. These
burst sources are found to be isotropically distributed on the sky
and are seen from a distance up to  50 parsecs.

\end{document}